\documentclass[aps,prc,preprint]{revtex4-1}
\usepackage[utf8]{inputenc}
\usepackage{graphicx}
\usepackage{amsmath, amssymb, bm, mathrsfs, gensymb}
\usepackage{enumitem}

\begin{document}
\title{Nucleon internal degrees of freedom and the uniqueness of the 
Gamow-Teller state}

\author{Naftali Auerbach} 
\email[]{auerbach@post.tau.ac.il}
\affiliation{School of Physics and Astronomy, Tel Aviv University, Tel Aviv 
69978, Israel.}

\date{\today}

\begin{abstract}
The Gamow-Teller strength in nuclei can be strongly affected by the internal 
degrees of freedom of the nucleon. It is demonstrated that this feature is 
unique to the Gamow-Teller. Excitation modes that involve spatial degrees of 
freedom are much less influenced by the internal excitations of nucleons. The 
fact that the observed Gamow-Teller strength is quenched by 30\%--40\% in all 
nuclei suggests that indeed this is due to the nucleon excitations in nuclei.
\end{abstract}


\maketitle

\section{Introduction}
Experimentally, for many years it is a well-established fact that the GT or M1 
strength is quenched \cite{Oste92, Gaav81}. In $\beta$ decay which involves the 
weak interaction or in charge-exchange reactions such as ($p,n$), ($^3$He,$t$), 
etc which are caused by the strong force, one observes strong quenching of the 
order 30--40\%. This is observed in all nuclei studied. The GT and M1 are 
unique in this respect. No such phenomenon is found for other types of 
excitations. We should, however, emphasize that this aspect of quenching of 
strength for other resonances, such as the monopole, quadrupole, etc was not 
yet studied in great detail. (For the isovector dipole, the strength is well 
accounted.) As for the observations of other types of strength, they are not 
detailed enough to determine whether there is significant strength missing.

Basically, two types of theories were introduced to discuss the quenching of GT 
and M1 strength in nuclei:
\begin{enumerate}[label=\Alph*.]
 \item The quenching of the GT strength in charge-exchange reactions refers to 
the main peaks where the GT strength is concentrated. So it was suggested that 
the remaining strength is fragmented and spread out at several tens of MeV 
above the main peaks \cite{Bert82,Klei85,Waka97}. The nuclear force (and 
especially the tensor component) causes a mixing of the one particle-one hole 
(1p-1h) components with 2p-2h excitations, causing the strength to be spread 
out (see \cite{Oste92,Bert82,Klei85,Waka97}). There are attempts to locate 
experimentally this fragmented strength, but it is quite difficult to reach a 
conclusive result. This mechanism could also affect other giant resonance, but 
no such effect was clearly observed so far.
 \item The other approach taken in the past to explain the missing GT, M1 
strength was to consider the influence of internal nucleon degrees of freedom 
that couple to the nucleus degrees of freedom and remove the strength to very 
high excitation energies, about 300 MeV above the ground state 
\cite{Brow81,Auer83,Zami82}.
\end{enumerate}
And then there is the possibility that both mechanisms A and B contribute to 
this puzzle of the missing GT strength. None of these possibilities can be 
ruled out at present. However, as we will see our considerations in this paper 
about the role of internal degrees of freedom makes the possibility B more 
plausible. The mechanisms described in A should be applied to the GT strength as 
well as to excitations involving the spatial coordinates. The understanding of 
the source of GT quenching has important consequences for nuclear experimental 
and theoretical physics. The quenching of the GT strength should be even more 
pronounced in double Gamow-Teller transition \cite{AuerZami89} and thus also in 
double-beta decay \cite{Enge17}. The two mechanisms will influence the 
two-neutrino double-beta decay calculations differently than the neutrinoless 
double-beta decay calculations \cite{Iach17}.

\section{Excitation of Giant resonances in Nuclei and the Role of Internal 
Nucleon degrees of Freedom}
Consider a one-body operator with space, spin, and isospin dependence acting on 
the nucleons in the nucleus:
\begin{equation}\label{first}
 \hat{O} = \sum_N \hat{O}_N + \sum_{q, N} \hat{O}_{q, N} \equiv \hat{O}_N + 
\hat{O}_q
\end{equation}
It is composed of two parts; the first sum referrers to operators acting on the 
nucleon degrees of freedom in the nucleus, the second sum involves 
single-particle operators acting on the internal degrees of freedom of each 
nucleon in the nucleus. We assume that the second sum contains only
transition operators from a nucleon to some baryonic resonances, and does 
not have diagonal parts. Let us now define a coherent nucleon particle-hole 
state:
\begin{equation}
 |\lambda \rangle = \frac{\hat{O}_N | 0 \rangle}{\langle 0 | \hat{O}_N^+ 
\hat{O}_N| 0 \rangle^{1/2}}
\end{equation} 
as well as a
\begin{equation}
 |\lambda^* \rangle = \frac{\hat{O}_q | 0 \rangle}{\langle 0 | \hat{O}_q^+ 
\hat{O}_q| 0 \rangle^{1/2}}
\end{equation} 
which represents a coherent sum of particle-hole excitations of the type $| 
N^{-1} N^*\rangle$ with definite parity, spin, and isospin; $N^*$ means excited 
nucleon states (resonances). 

Considering the mixed state:
\begin{equation}
 | \Phi \rangle = | \lambda \rangle + \alpha | \lambda^* \rangle,
\end{equation} 
with $\alpha$ being the mixing amplitude, we now evaluate the transition 
strength between the ground state $| 0 \rangle$ and the state $| \Phi \rangle$ 
for the operator $\hat{O}$
\begin{equation}
 |\langle \Phi | \hat{O} | 0 \rangle|^2 = | \langle \lambda| \hat{O}_N | 0 
\rangle + \alpha \langle \lambda^* | \hat{O}_q | 0 \rangle |^2
\end{equation}
We assume that the second term is small enough compared to the first one so we 
can neglect terms proportional to $\alpha^2$. Consequently,
\begin{equation}
 | \langle \Phi | \hat{O} | 0 \rangle |^2 = |\langle \lambda | \hat{O}_N | 0 
\rangle|^2 \left[ 1 + 2\alpha \frac{\langle \lambda^*| \hat{O}_q | 0 \rangle }{ 
\langle \lambda |\hat{O}_N | 0 \rangle } \right].
\end{equation}
The expression
\begin{equation}
 \gamma = 2\alpha \frac{ \langle \lambda^*| \hat{O}_q | 0 \rangle}{\langle 
\lambda | \hat{O}_N | 0 \rangle}
\end{equation} 
is the ``quenching factor'', although it could also be an enhancement depending 
on the sign of $\alpha$. It represents the contribution of the internal 
excitations to the strength at low energies.

The mixing amplitude is
\begin{equation}
 \alpha = 2 \frac{ \langle \lambda| \sum V_{NN^*} | \lambda^* 
\rangle}{E_{\lambda^*}},
\end{equation}
where $V_{NN^*}$ is a two-body transition potential for $NN \rightarrow NN^*$ 
and $E_{N^*}$ is the excitation energy of $ | \lambda^* \rangle$ with respect 
to the g.s. and is approximately equal to the mass of the $N^*$ relative to the 
nucleon mass, $E_{\lambda^*} \simeq m_{N^*} - m_N$.

The excitation energy $\hbar \omega$ in the denominator was neglected because 
$\hbar \omega  << E_{N^*}$. Following several calculations in the past 
\cite{Brow81} (in particular those involving the $\Delta_{33}$ resonance) one 
can write:
\begin{equation}
 \langle \lambda | \sum V_{NN*}| \lambda^* \rangle \simeq \langle \lambda 
| \sum V_{NN}| \lambda \rangle \frac{\langle \lambda^*| \hat{O}_q| 0 
\rangle}{\langle 
\lambda | \hat{O}_N | 0 \rangle} G,
\end{equation} 
where $G$ is the ratio of the coupling constant:
\begin{equation}
 G = \frac{g_{mNN^*}}{g_{mNN}},
\end{equation} 
$m$ stands for the meson exchange, mostly the pion- $\pi$.

We can write $\gamma$ in the form:
\begin{equation}\label{gamma}
 \gamma = 2\frac{ \langle \lambda | \sum V_{NN} | \lambda \rangle 
}{E_{\lambda^*}} \frac{\langle \lambda^*| \hat{O}_q | 0 \rangle^2}{\langle 
\lambda^*| \hat{O}_N | 0 \rangle^2}G.
\end{equation}
The matrix element squared $\langle \lambda^*| \hat{O}_q | 0 \rangle^2$ can be 
estimated in the following way
\begin{equation}
 \langle \lambda^*| \hat{O}_q | 0 \rangle^2 = Z\delta_p + N \delta_n
\end{equation} 
where $Z$ and $N$ denote the number of protons and neutrons in the nucleus. 
$\delta_p$ and $\delta_n$ are $\langle p |\hat{O}_q| N^*_{1/2}\rangle$ and 
$\langle n |\hat{O}_q| N^*_{-1/2}\rangle$ corresponding the transition 
strengths between the proton (neutron) and the $N^*$ component with $T_z = -1/2$ 
and the $T_z = 1/2$.

For the sake of an estimate, it is reasonable to assume \cite{Auer83} that 
$\delta_p = \delta_n = \delta$. We see that the expression of $\gamma$ will be 
enhanced by the factor $A$, the number of nucleons in the nucleus. The ratio of 
the coupling constant G is of the order of $1$ \cite{Brow81} and the energy 
$E_{\lambda^*}$ is several hundred MeV (about 300 MeV for the $\Delta_{33}$ 
resonance, 600 MeV for $D'_{13}$ resonance, 500 MeV for the Roper resonance, 
etc).

The matrix element $ \langle \lambda| \sum V_{NN} | \lambda \rangle $ can be 
estimated by realizing that this is a shift of the excitation due to the 
particle-hole interaction \cite{Zami82, AuerKlei83}. Such matrix element is 
typically of the order of $\hbar \omega = x 41 A^{-1/3}$ MeV, where $|x|$ is 
about 1--2 depending on the type of nuclear excitation and the nucleus. For 
isoscalar excitations, $x$ is negative because the particle-hole interaction is 
attractive but for isovector excitation it is repulsive and $x$ is positive.

Let us now estimate the ratio
\begin{equation}\label{f}
 f = \frac{\delta}{\langle \lambda | \hat{O}_N | 0 \rangle^2}
\end{equation}
Here we must distinguish two types of nuclear excitations. The ones that depend 
on the spatial coordinate $\bm{r}$ and those that do not. Many of the 
collective nuclear excitations are $\bm{r}$-dependent. The best example is the 
isovector dipole resonance. The corresponding operator exciting the dipole 
resonance is $ \hat{O}(\text{dip}) = \sum_i r_i Y_1(\theta_i) \bm{t}_i.$
But there are more; isovector monopole, quadrupole with the operators
$ \sum_i r_i^2 \bm{t}_i, \sum_i r_i^2 Y_2(\theta_i) \bm{t}_i$, etc. One can 
also introduce the spin into these operators. As for example the spin-isovector 
dipole $\sum_i \sigma_i r_i Y_1(\theta_i) \bm t_i$, spin-isovector monopole 
$\sum \sigma_i r_i^2 \bm t_i$, etc.

There are also the excitations that do not involve spatial coordinates, only 
spin-isospin operators, the Gamow-Teller (GT) and M1 with operators
\begin{equation}
 \hat{O}_N = \sum_i \sigma_i \bm t_i.
\end{equation} 

\section{Transition operators with spatial dependence}
When we estimate the ratio $f$ (in Eq. (\ref{f})) for excitations involving the 
radial coordinate mentioned above we understand  that the $\hat{O}_q$ operator 
will involve also the same type of spatial coordinate but for quarks inside the 
nucleon, thus $\bm r_N$ \cite{Auer83}. Therefore, the ratio $f$ will be 
proportional to $(r_N/R)^n$, that is the ratio of the radius of the nucleon 
(or the $N^*$ resonance) and the radius of the nucleus-$R$ to some power $n 
\geq 2$, depending on the type of excitation one considers. For example for a 
dipole $n=2$, for a monopole $n=4$, etc. Clearly, this ratio is small. It means 
that the ratio $f$ will be reduced by a large factor for a dipole, quadrupole, 
monopole, etc. 

In the past, we estimated this factor \cite{Auer83} for the dipole 
taking into account the coupling to the $D'_{13}$ resonance and using the 
experimental $\sigma_{-1} \simeq 30 \mu b$ cross sections for the $N \rightarrow 
D'_{13}$ transition \cite{Kell80} and for the nuclear dipole from Ref. 
\cite{Berm75}. We found the reduction factor $f$ to be about 1/2000 and 
the quenching factor $\gamma$ to be $4 \times 10^{-3}$ for $^{90}$Zr and $1.5 
\times 10^{-3}$ for $^{208}$Pb.

\section{Transition operators with no spatial dependence}
As opposed to the transition operators which depend on the spatial coordinate 
we will now deal with simple operators that do not depend on $\bm r$. Among 
the simple ones there is one outstanding operator, the Gamow-Teller operator 
(see for example \cite{Oste92})
\begin{equation}
 \hat{O}_{GT} = \sum_i \sigma(i) t_\mu(i); \quad \mu = 0, \pm 1.
\end{equation} 
We will concentrate on the $\mu = -1$ component of this operator. The GT 
excitation of the nucleon is the $\Delta_{33}$ resonance with a mass of 1232 
MeV. Eq. (\ref{first}) will be written now as
\begin{equation}
 \hat{O}_{GT} = \hat{O}_{N} + \hat{O}_{\Delta},
\end{equation}
where 
\begin{equation}
 \hat{O}_{N} = \sum_i \sigma (i) t_- (i),
\end{equation}
and
\begin{equation}
 \hat{O}_{\Delta} = \sum_i S(i) T_- (i)
\end{equation} 
is the nucleon to $\Delta$ part of the GT operator \cite{Brow81,Zami82}.

Let us now use Eq. (\ref{gamma}) to estimate the quenching factor $\gamma$, for 
$|\lambda \rangle = |GT \rangle$ (For simplicity we will deal with $N > 
Z$ nuclei and with cases when the $GT_+$ excitations are blocked by the Pauli 
principle.). The matrix element $\langle GT| \sum V_{NN} | GT \rangle$ is the 
shift of the GT state from the unperturbed position. In the nucleus $^{208}$Pb 
this shift is about 10 MeV. The energy $E_{\lambda^*}$ in Eq. (\ref{gamma}) is 
the position of the $\Delta$ resonance above the nucleon, that is about 300 
MeV. 
The transition matrix element squared $|\langle 0| \hat{O} |GT \rangle |^2$ of 
the particle-hole state in the nucleus with a nucleon converted into a 
$\Delta_{33}$ is according to a quark model \cite{Brow81} equal to 
$\frac{32}{25}A$, the matrix element squared $\langle GT | \hat{O}_N| 0 
\rangle^2$ for the GT is 
$N-Z$ (the strength of the GT excitation) and the coupling constant $G = 4/3$. 
Putting all these numbers into Eq. (\ref{gamma}), we obtain $\gamma = 0.53$ for 
$^{208}$Pb, and $\gamma = 0.51$ for $^{90}$Zr. 
Thus for the GT strength, the quenching factor is of the order of one. Our 
estimates are not precise and the numbers we present for $\gamma$ 
are approximate, however one sees a qualitative change when one compares the 
quenching factor for the excitations involving only spin-isospin operators to 
the ones that involve spatial coordinates.

\section{Conclusion}
Our analysis shows that the considerable quenching of strength due to internal 
excitation of the nucleon occurs uniquely for the GT or 
M1. Experimentally the clear and systematic quenching of strength is 
found only for the GT or M1 and not for the multipole $r$-dependent 
resonances. This suggests that the GT quenching is to a large extent due to the 
mixing with the $\Delta$ particle-N hole configurations. 
Often a question is asked if the GT or M1 strength is quenched, does this apply 
to other types of excitations involving the spin. For example, are the so-called 
``first forbidden'' $\beta$ decay transitions also reduced? The operators 
relevant to such transition are 
of the type $\bm{r} \sigma \bm {t}$ or $\bm{r}^2 \sigma \bm {t}$
\cite{BMbook,AuerKlei84}, isovector dipole resonance, and isovector spin 
monopole, respectively. Some of these were observed, the spin dipole 
\cite{Kras99} and spin isovector monopole \cite{Miki12}. These observations are 
not detailed enough to determine whether there is significant strength missing. 
If the quenching of strength in the GT resonance is mostly due to the mixing 
with $\Delta_{33}$, then in the view of our discussion, the above spin 
transitions which contain the coordinate $\bm r$ will not be considerably 
quenched. Obviously, there is still a lot of experimental work to be done to 
determine the total strength of $r$-dependent excitations.

In summary, the GT strength is unique when considering the influence of 
internal degrees of freedom of the nucleon. The reason is that the GT and M1 
excitation do not involve a spatial coordinate. Many of the nucleons in the 
nucleus occupy both the spin up and down states and the Pauli principle limits 
therefore the nuclear GT or M1 transitions. The second main reason is that the 
internal excitations of the quarks are spatially confined. This confinement 
limits the amplitudes of the spatial vibrations when compared to the analogous 
nuclear vibrations.

\end{document}